\documentclass[11pt]{article}

\usepackage{graphicx}	
\usepackage{amsmath} 	
\usepackage{amssymb}
\usepackage{url}
\usepackage{authblk}
\usepackage{color}
\usepackage{subfig}
\usepackage{fullpage}

\newcommand{\Eref}[1]{(\ref{eq:#1})}

\newcommand{\Fref}[1]{Figure \ref{fig:#1}}
\newcommand\labelthis{\addtocounter{equation}{1}\tag{\theequation}}

\newcommand{\epsA}[0]{\varepsilon_{\mathrm{A}}}
\newcommand{\epsB}[0]{\varepsilon_{\mathrm{B}}}
\newcommand{\epsO}[0]{\varepsilon_{0}}
\newcommand{\epsR}[0]{\varepsilon_{\mathrm{r}}}
\newcommand{\epsL}[0]{\varepsilon_{\mathrm{L}}}
\newcommand{\muA}[0]{\mu_{\mathrm{A}}}
\newcommand{\muB}[0]{\mu_{\mathrm{B}}}
\newcommand{\muO}[0]{\mu_{0}}
\newcommand{\muL}[0]{\mu_{\mathrm{L}}}
\newcommand{\vE}[0]{\mathbf{E}}
\newcommand{\vH}[0]{\mathbf{H}}
\newcommand{\vB}[0]{\mathbf{B}}
\newcommand{\vD}[0]{\mathbf{D}}
\newcommand{\vU}[0]{\mathbf{U}}

\newcommand{\aL}[0]{\alpha_{\mathrm{L}}}
\newcommand{\wP}[0]{\omega_{\mathrm{p}}}
\newcommand{\wO}[0]{\omega_{0}}
\newcommand{\chiA}[0]{\chi_{\mathrm{A}}}
\newcommand{\chiB}[0]{\chi_{\mathrm{B}}}
\newcommand{\chiL}[0]{\chi_{\mathrm{L}}}
\newcommand{\Mod}[1]{\lvert #1 \rvert}
\newcommand{\vp}[0]{v_{\mathrm{p}}}

\newcommand{\abs}[1]{\left\lvert #1 \right\rvert}

\newtheorem{remark}{\bf Remark}[section]

\begin{document}

\title{\sc Frequency-dependent impedance and surface waves
 on the boundary of a stratified dielectric medium}


\author[1]{Kirill Cherednichenko}
\author[1]{William Graham}
\affil[1]{Department of Mathematical Sciences, University of Bath, Claverton Down, Bath, BA2 7AY, United Kingdom}



\maketitle



\begin{abstract}
We analyse waves propagating along the interface between half-spaces filled with a perfect dielectric and a Lorentz material. We show that the corresponding interface condition leads to a generalisation of the classical Leontovich condition on the boundary of a dielectric half-space. We study when this condition supports propagation of (dispersive) surface waves.  We derive the related dispersion relation for waves along the boundary of a stratified half-space and determine the relationship between the loss parameter, frequency, and wavenumber for which interfacial waves exist.

\end{abstract}









\section{Introduction} 
\label{sec:Intro}

The question of what boundary conditions are satisfied by an electromagnetic field 
 has been touched upon in the mathematics literature a number of times since the work by Leontovich in the early 1940's, see \cite{Leontovich}, in relation to wave propagation near the surface of the earth. 
The related setup of an interface between the free half-space (or its curved analogue) and an electromagnetic medium with a large refractive index is well studied from the physics, analysis, and numerics perspectives, and does not present a challenge with the modern computational power available, see {\it e.g.} \cite{Senior_Volakis}. However, 
the question of what form the Leontovich condition takes when neither of the two media is nearly perfectly conducting is worth exploring, as replacing the interface conditions by ``effective'' conditions on the boundary of a dielectric medium can lead to a reduction in computational demand, and the information about the dispersion relation for such waves may be exploited for the design of new transmission devices. 

The wider subject of the derivation of effective conditions along an interface between electromagnetic media has been studied extensively during the last few decades, see {\it e.g.} \cite{Slepyan1}, \cite{Slepyan2}, although in many situations the analysis remains outstanding, especially when the media in contact are heterogeneous or frequency-dispersive. 
As often noted in the existing literature, interfacial and surface waves are often amenable to a direct analysis in view of the simplified geometry, as the solutions admit a separation 
into oscillations along the surface and exponential decay away from it.  This observation prompts us to try and obtain closed-form dispersion relations for the surface of contact between a dielectric and a general Lorentz medium, at least in the case of a flat interface.  
The related analysis should admit further generalisation towards the curved case when the typical radius of curvature is assumed small compared to the wavelength, see \cite{BabichKuznetsov} for the case of the classical Leontovich condition, which we later explore as a limit of a more general condition. 

Irrespective of the question of effective boundary conditions, the subject of electromagnenic surface wave propagation, {\it i.e.} wave motion localised the interface between two electromagnetic media, has received a great amount of attention in the physics and applied mathematics literature since mid-20th century, see \cite{PML} for a detailed overview of the subject and extensive bibliography. In particular the case of stratified dielectric media in contact with a homogeneous dielectric was introduced in \cite{YYH}, with a number of papers discussing its applications following, see {\it e.g.} \cite{YYC}, \cite{Robertson_May}, \cite{Robertson}. The case when the homogeneous dielectric half space is replaced by a frequency dispersive (say, Lorentz) medium, remains largely unexplored from the point of view of the analysis of interfacial modes.
 
In the present paper we combine the above two open directions and discuss the case of a two-component stratified dielectric in contact with a Lorentz medium. In particular, we derive the corresponding version of the Leontovich condition, and analyse the associated waves along the contact surface.
Assuming that the dielectric properties are periodic in the direction orthogonal to the contact surface allows us to consider a non-homogeneous case, while retaining the ability to carry out an explicit analysis by virtue of the Floquet theory, see Section \ref{Maxwell_section}. Using an effective boundary condition on the surface of contact with the complementary half-space occupied by the Lorentz medium, obtained in Section \ref{sec:LowLor} by imposing the condition of decay of the wave amplitude away from the surface into this half-space, in Section \ref{sec:EqHSProb} we derive a dispersion relation between the wave-number and frequency for waves propagating along the interface and decaying in amplitude away from it. In Section \ref{sec:EqHSProb} we also investigate how the wavenumber-frequency pairs depend on the loss parameter of Lorentz half-space and how the dispersion diagrams for the lossless case depend on the ratio between the plasma and resonant frequencies. Finally, in Section \ref{GeneralToClassical} we discuss the relation between our effective condition and the standard Leontovich condition.






\section{Problem setup} \label{sec:Setup}

\subsection{Maxwell system} \label{sec:Maxwell_Setup}

The Maxwell equations of electromagnetism are, see {\it e.g.} \cite{Jackson}:
\begin{equation} 
\label{eq:Maxwell} 
\nabla \wedge \vE 
= -\dfrac{\partial \vB}{\partial t}, \quad\quad\quad
\nabla \wedge \vH 
= \dfrac{\partial \vD}{\partial t},
\end{equation}
where $t$ is time, $\vE$, $\vH$ are the electric and magnetic field, $\vD$, $\vB$ are the electric displacement and magnetic flux density (or ``magnetic induction''), and $\wedge$ denoted the standard 3-vector cross product: $\nabla\wedge\mathbf{A}:=\mathrm{curl}\mathbf{A}$ for a vector field $\mathbf{A}.$ 

The displacement $\vD$ and induction $\vB$ are related to the electric $\vE$ and magnetic $\vH$ fields via the constitutive laws
\begin{equation}
\vD=\varepsilon\vE,\quad\vB=\mu\vH,
\label{constitutive_laws}
\end{equation}
where the permittivity $\varepsilon$ (``electric permittivity'', or ``dielectric constant'') and $\mu$ (``magnetic permeability'') are material parameters, which may depend on the frequency $\omega,$ see Section \ref{Lorentz_section}. In what follows we denote by $\epsO$ (respectively $\muO$)  the permittivity (respectively, permeability) of free space (``vacuum").

We consider the system (\ref{eq:Maxwell}) either in a half-space $\{\mathbf{x}=(x_1, x_2, x_3)\in\mathbb{R}: x_{3}>0\},$ with a boundary condition at $\{x_{3}=0\}$ and a decay condition as $x_3\to\infty$, or in the full space ${\mathbb R}^3$ with an interface condition between two materials at $\{x_{3}=0\}$ and decay conditions as $x_3\to\pm\infty.$ 
Without loss of generality, we seek waves propagating in the $x_1$ direction on the $\{x_{3}=0\}$ surface (or interface), {\it i.e.} solutions to (\ref{eq:Maxwell}) of the form
\begin{equation}
\label{eq:SolForm} 
\vE (\mathbf{x})= \begin{pmatrix} E_{1}(x_3) \\[0.4em] E_{2}(x_3) \\[0.4em] E_{3}(x_3) \end{pmatrix} \exp\bigl({\rm i}(k x_{1} -\omega t)\bigr), \quad\quad
\vB (\mathbf{x})= \begin{pmatrix} B_{1}(x_3) \\[0.4em] B_{2}(x_3) \\[0.4em] B_{3}(x_3) \end{pmatrix} \exp\bigl({\rm i}(k x_{1} -\omega t)\bigr),\quad x_1,x_3\in{\mathbb R},
\end{equation}
where $k$ is the so-called wavenumber, which has the dimensions of inverse length. In what follows, we choose to work with the amplitude components 
 continuous across the interface: $E_{1},$ $E_{2},$ $D_{3},$ $H_{1},$ $H_{2},$ $B_{3}.$ 

\subsection{Lorentz materials}
\label{Lorentz_section}
In the Lorentz oscillator model for the optical properties of materials \cite{AlmogEtAl, Nussenzveig, Rosenfeld}, 
electrons
are considered bound to the nuclei, and the binding force interaction is represented by a ``mass-on-a-spring'' system under the assumption that the nucleus is far more massive than the electron and hence does not change its position. 
A damping term is introduced, to account for the inherent loss of energy as the (charged) electron accelerates, and the system is subject to a driving force of the same frequency as an incident electromagnetic radiation. 
The result of this construction 
is an $\omega$-dependent relative permittivity $\epsL(\omega):=\varepsilon/\varepsilon_0:$
\begin{equation}
\label{eL_er}
\epsL(\omega)=\epsR(\omega)+{\rm i}\frac{\sigma(\omega)}{\omega},\qquad\omega>0,
\end{equation}
\begin{equation}
\label{eq:lorEps} 
\epsR(\omega): = 1 + \frac{\wP^2(\wO^{2}-\omega^{2})}{(\wO^{2}-\omega^{2})^{2}+(\omega\gamma)^{2}},\qquad
 \sigma(\omega):=-\frac{(\wP\omega)^2\gamma}{(\wO^{2}-\omega^{2})^{2}+(\omega\gamma)^{2}},
\end{equation}
where $\wP$ and $\wO$ are the so-called plasma and resonant frequencies 
and $\gamma$ is the loss factor, 
all of which are material constants with the dimension of frequency.
Note that $\epsR(\omega)$ and $\sigma(\omega)$ are real-valued whenever $\omega$ is real. 
In what follows we assume that $\mu=\mu_0,$ although the Lorentz theory can also be used to obtain the relative permeability $\muL=\mu/\mu_0$ as a function of $\omega$, for any imperfectly conducting material that admits polarisation by an external magnetic field.
	
In addition, we note that a similar theory for the dependence of permittivity and permeability on $\omega$ has been developed for metals \cite[Chapter 7]{Jackson}, which yields the form \Eref{lorEps} with 
$\wO=0.$ The convention of expressing $\gamma$ in terms of the mean travel time between electron collisions is also commonly adopted in this case.


\section{Maxwell system in a homogeneous space} 
\label{sec:HomMaxwell}

In this section we discuss solutions to two auxiliary problems associated with the system (\ref{eq:Maxwell})--(\ref{constitutive_laws}) that describe the propagation of waves in a homogeneous medium, either dielectric (Section \ref{sec:SingLayMax}) or Lorentz (Section \ref{sec:LowLor}).

\subsection{Classical solution in free space} 
\label{sec:SingLayMax}
In view of our objective the study the Maxwell system in a stratified half-space in Section \ref{Maxwell_section}, we  consider a homogeneous, isotropic dielectric material of permittivity $\varepsilon$ and permeability $\mu$ occupying the region $\{\mathbf{x}\in\mathbb{R}:x_{3}\in(0,a)\}$ for some constant $a\in\mathbb{R},$ which will represent the thickness of individual layers in Section \ref{Maxwell_section}.
Using the ansatz \Eref{SolForm} yields two systems 
\begin{equation}
\left\{\begin{array}{lll}
E_{1}'=
 i\omega\mu H_{2} + \dfrac{{\rm i}k}{\varepsilon}D_{3},\\[0.7em]
H_{2}'= i\omega\varepsilon E_{1},\\[0.9em]
-\omega D_{3} = kH_{2},\end{array}\right.\qquad\qquad
\left\{\begin{array}{lll}
E_{2}'= -{\rm i}\omega\mu H_{1}, \\[0.7em]
H_{1}'= \dfrac{{\rm i}k}{\mu}B_{3} - {\rm i}\omega\varepsilon E_{2}, \\[0.9em]
\quad \omega B_{3} = kE_{2},
\end{array}\right.
\end{equation}
each consisting of two differential equations and one algebraic equation. 
Our analysis henceforth focuses on the transverse electric (TE) polarisation\footnote{Solutions to the transverse electric system, or ``polarisation" are also referred to as the ``electric waves''. Correspondingly, the transverse magnetic solutions is sometimes referred to as the ``magnetic waves'' in 
	the literature, see {\it e.g.} \cite{BabichKuznetsov}.}
	 involving the field components $E_{1}, H_{2}, D_{3}$, which can be expressed in matrix form as
\begin{equation}
\label{eq:TESys}
\begin{pmatrix} E_{1}' \\[0.4em] H_{2}' \end{pmatrix} = 
\begin{pmatrix} 
0	& -\dfrac{{\rm i}\alpha^{2}}{\omega\varepsilon} \\[0.7em]
{\rm i}\omega\varepsilon	& 0 
\end{pmatrix}
\begin{pmatrix} E_{1} \\[0.4em] H_{2} \end{pmatrix},\qquad \alpha^{2} := k^{2}-\omega^{2}\mu\varepsilon.
\end{equation}
The above $(2\times2)$-system for $\vU:=(E_1, H_2)^\top$ 
 is solved by diagonalising the matrix $A:$ 
\begin{equation*}
\label{eq:FormSolU}
\vU(x_{3})= \exp(A x_{3})\vU(0) 
			= \begin{pmatrix} \cosh(\alpha x_{3}) & -\dfrac{{\rm i}\alpha}{\omega\varepsilon}\sinh(\alpha x_{3}) 
			\\[0.8em]
\dfrac{{\rm i}\omega\varepsilon}{\alpha}\sinh(\alpha x_{3}) & \cosh(\alpha x_{3}) \end{pmatrix} \vU(0),\quad\ \  x_3\in(0,a).
\end{equation*}
We shall use this form for the solution when considering the Maxwell system in a stratified material in Section  \ref{Maxwell_section}. 
For completeness, we note that calculations for the transverse magnetic (TM) polarisation, analogous to those above for the TE case, result in a system of the form (\ref{eq:TESys}), with $\vU$ replaced by 
$\mathbf{V}:=\left(H_{1}, E_{2}\right)^\top$ and $\varepsilon$ replaced by $-\mu.$

\subsection{Decaying solution in Lorentz half-space} 
\label{sec:LowLor}

As preparation for considering the full-space problem, consider \Eref{Maxwell} in the half-space $\{x_{3}<0\}$, occupied by a Lorentz material with permittivity described in (\ref{eL_er})--\Eref{lorEps}.
We 
 impose a decay condition away from the boundary, seeking solutions that tend to zero as $x_{3}\rightarrow-\infty$, which yields 
\begin{equation*} 
E_{1}(x_{3}) = C\frac{{\rm i}\aL}{\omega\epsL\epsO}\exp(\aL x_{3}),\qquad
H_{2}(x_{3}) = C\exp(\aL x_{3}),\qquad x_3<0,\qquad\quad C\in{\mathbb C},
\end{equation*}
\begin{equation}
\aL(k,\omega):=\sqrt{k^{2}-\omega^{2}\epsO\epsL\muO\muL},\qquad \arg(\aL)\in\left(-\pi/2,\pi/2\right].
\label{alpha_L} 
\end{equation}
In particular, the following condition at $x_{3}=0$ is satisfied:
\begin{equation}
\label{eq:GeneralLeontovich}
E_{1}(0) =-\frac{{\rm i}\aL}{\omega\epsL\epsO}H_{2}(0),
\end{equation}
which is similar to the classical Leontovich impedance boundary condition 
\cite{Senior}. 
The quantity $\aL/(\omega\epsL\epsO)$ has physical dimensions of Ohms and plays a r\^{o}le analogous to the  impedance in the classical condition, and in what follows we refer to it as the generalised impedance.
In Section \ref{GLC}
we show how the condition (\ref{eq:GeneralLeontovich}) is used in the analysis of boundary-value problems,
and in Section \ref{GeneralToClassical} we explore its relation 
to the classical Leontovich condition. 

In conclusion of this section we note that considering similarly the TM polarisation yields the boundary condition
\begin{equation*}	
\label{eq:GLC_TM}
	E_{2}(0) = \frac{{\rm i}\omega\muO\muL}{\aL} H_{1}(0),
\end{equation*} 
where $\muL$ is the $\omega$-dependent magnetic permittivity.
\section{Maxwell system for stratified media}
\label{Maxwell_section}

Consider the upper half-space $\{x_{3}>0\}$ occupied by a stratified dielectric, {\it i.e.} a medium consisting of alternating layers of materials A and B, parallel to the $x_{3}$-plane, with permittivities $\epsA$ and $\epsB$, and permeabilities $\muA$ and $\muB$ respectively. 
We denote the period of the structure by $d,$ so that A-layers have thickness $dh$ and B-layers have thickness $d(1-h)$ for some $h\in(0,1)$. 
We wish to obtain solutions of the Maxwell system \Eref{Maxwell} that decay as $x_{3}\rightarrow\infty,$ subject to interface conditions (continuity of the fields) at $x_{3}=dh,d,d(1+h),2d,\dots.$ 

\subsection{Floquet analysis of arbitrary whole-space solutions}
\label{Floquet_section}

In this section we review results on 
matrix differential equations and Floquet theory, which we shall invoke when solving the Maxwell system in the region $\{x_3>0\}.$ These are specific for the polarisation and geometry we consider in the present paper but can be developed for other setups, based on the general analytical approach, see {\it e.g.} \cite{Eastham}. 


In the stratified half-space the Maxwell system has the form
\begin{equation} 
\label{eq:Flo1}
\vU'(x)=A(x)\vU(x),\qquad x>0,
\end{equation}
where we write $x$ instead of $x_3$ for brevity, $\vU$ is a 2-vector, and $A$ a piecewise-constant $(2\times2)$-matrix
\begin{align*}
A(x) = \begin{cases}
A_{1}, &0<x<dh	,\\[0.3em]
A_{2}, &dh<x<d, \end{cases}
\end{align*}
extended $d$-periodically. The matrices $A_{1}$ and $A_{2}$ have the form of the matrix in \Eref{TESys}, with the general $\varepsilon$ and $\mu$ replaced by the constants specific to the materials A and B. Each of them has two distinct eigenvalues
and hence $A_j = T_j\Lambda_jT_j^{-1},$ $j=1,2,$  for diagonal $\Lambda_j$ and transformation matrices $T_j,$ whose columns are eigenvectors of $A_j.$ 
It follows that
\begin{align*}
\vU(x) = \begin{cases}
T_{1}\exp(\Lambda_{1}x)T_{1}^{-1}\vU(0), &0<x\leq dh, \\[0.4em]
T_{2}\exp\bigl(\Lambda_{2}(x-dh)\bigr)T_{2}^{-1}T_{1}\exp(\Lambda_{1}dh)T_{1}^{-1}\vU(0), &dh<x\leq d, \\ \end{cases}
\end{align*}
by solving (\ref{eq:Flo1}) in each layer 
and using the continuity condition at $\{x_{3}=dh\}.$ 
In particular,
\begin{equation}
\label{eq:EndpointEqns} 
\begin{aligned} 
&\vU(dh) = T_{1}\exp(\Lambda_{1} dh)T_{1}^{-1}\vU(0), \\[0.3em]
&\vU(d)= T_{2}\exp\bigl(\Lambda_{2}d(1-h)\bigr)T_{2}^{-1}T_{1}\exp(\Lambda_{1}dh)T_{1}^{-1}\vU(0).
\end{aligned} 
\end{equation}

The general theory systems of linear ODEs implies the existence of an invertible matrix function $\Phi$ (``fundamental matrix'') such that for any solution $\vU$ to (\ref{eq:Flo1}) one has
$\vU(x)=\Phi(x)\Phi(0)^{-1}\mathbf{U}(0),$ $x>0.$
Taking $\Phi(\cdot)\Phi(0)^{-1}$ instead of $\Phi(\cdot)$ if necessary shows that one can always choose $\Phi$ so that $\Phi(0)=I$ ({\it i.e.} $\Phi$ is the ``canonical fundamental matrix''), which we do henceforth. 

In what follows we show that there exists a fundamental matrix $\widetilde{\Phi}$ for  
\Eref{Flo1} of the form
\begin{equation}
\widetilde{\Phi}(x)=\widetilde{\Psi}(x)\mathrm{diag}\bigl\{\exp(\widetilde{\lambda}_{1}x), \exp(\widetilde{\lambda}_{2}x)\bigr\},\quad x>0, 
\label{Phi_tilde_form}
\end{equation}
where $\widetilde{\lambda}_1,$ $\widetilde{\lambda}_2\in{\mathbb C},$ and $\widetilde{\Psi}(x)$ is a $d$-periodic matrix. 

The matrix ({\it cf.} (\ref{eq:EndpointEqns}))
\begin{align*}
\Phi(d) &= T_{2}\exp\bigl(\Lambda_{2}d(1-h)\bigr)T_{2}^{-1}T_{1}\exp(\Lambda_{1}dh)T_{1}^{-1}
\end{align*}
is referred to as the monodromy matrix  (or ``transfer matrix"). We write 
 $\Phi(d)={\mathbb T}\,\mathrm{diag}\left(\lambda_{1},\lambda_{2}\right){\mathbb T}^{-1},$ where 
$\lambda_{1},\lambda_{2}$ are the eigenvalues of $\Phi(d),$ so that ${\mathbb T}$ is a matrix whose columns are the corresponding eigenvectors of $\Phi,$ and define the matrix function $\Psi$ as follows:
\begin{equation*}
\Psi(x):=\Phi(x){\mathbb T}\,\mathrm{diag}\biggl\{\exp\left(-\frac{x}{d}\ln\lambda_{1}\right), \exp\left(-\frac{x}{d}\ln\lambda_{2}\right)\biggr\}{\mathbb T}^{-1},\qquad x\in(0,d],
\label{Psi_form}
\end{equation*}
where we use the principal value of the logarithm. 

Denote by $\widehat{\Psi}$ the $d$-periodic extension of $\Psi$ to $(0,\infty).$ We claim that the function
\begin{equation*} 
\label{eq:ClaimForm}
\widehat{\Phi}(x):=\widehat{\Psi}(x){\mathbb T}\,\mathrm{diag}\biggl\{\exp\left(\dfrac{x}{d}\ln\lambda_{1}\right),\exp\left(\dfrac{x}{d}\ln\lambda_{2}\right)\biggr\}{\mathbb T}^{-1},\qquad x>0,
\end{equation*}
coinsides with $\Phi.$
To see this, note first that 
$\widehat{\Phi}'=A\widehat{\Phi}$
everywhere except at $d, 2d, 3d,\dots,$ and is continuous at these points, since
\begin{equation*}
\widehat{\Psi}(d)=\Psi(d)=\Phi(d){\mathbb T}\,\mathrm{diag}\bigl\{\exp\bigl(-\ln\lambda_{1}\bigr), \exp\bigl(-\ln\lambda_{2}\bigr)\bigr\}{\mathbb T}^{-1} 
	= \Phi(d)\Phi(d)^{-1}=I=\widehat{\Psi}(0),
\end{equation*}
and hence $\widehat{\Psi}$ is continuous. Since $\widehat{\Phi}(0)=\widehat{\Psi}(0)=I=\Phi(0),$ one has $\widehat{\Phi}=\Phi$ by the uniqueness theorem for (\ref{eq:Flo1}), see {\it e.g.} \cite{Coddington_Levinson}.

Multiplying both sides of (\ref{eq:ClaimForm})
 by ${\mathbb T},$ we find that the fundamental matrix (\ref{Phi_tilde_form}) is given by
\begin{equation*}
\widetilde{\Phi}(x) :=\Phi(x){\mathbb T} 
     =\widehat{\Psi}(x){\mathbb T}\,\mathrm{diag}\biggl\{\exp\left(\frac{x}{d}\ln\lambda_{1}\right), 
				\exp\left(\frac{x}{d}\ln\lambda_{2}\right)\biggr\},
				\quad x>0.
\end{equation*}
where 
$\widetilde{\Psi}:=\widehat{\Psi}{\mathbb T}$ is $d$-periodic by the construction of $\widehat{\Psi}.$

An immediate consequence of the above is that any solution $\vU$ to \Eref{Flo1} has the form
\begin{equation}
\label{eq:solForm}
\vU(x) 	= \widetilde{\Phi}(x)
{\mathbf{C}}
		={c}_{1}
		\exp\left(\dfrac{x}{d}\ln\lambda_{1}\right)\widetilde{\Psi}_{1}(x) 
		+ 
		{c}_{2}\exp\left(\dfrac{x}{d}\ln\lambda_{2}\right)\widetilde{\Psi}_{2}(x),\quad x>0,
\end{equation}
with a constant vector $\mathbf{C}=({c}_{1}, {c}_{2})^\top$ and $\widetilde{\Psi}_j,$ $j=1,2,$ denoting the $j^{\mathrm{th}}$-column of the matrix $\widetilde{\Psi}.$

\subsection{Decaying solution in the stratified half-space} 
\label{sec:UppHSSol}


As the matrices $A_1,$ $A_2$ are traceless, {\it cf.} \Eref{TESys}, one has
 $\lambda_{1}\lambda_{2}=1,$
 which gives two possible cases:
	$\lambda_{1},\lambda_{2}=\lambda_1^{-1}\in \mathbb{R},$ $\vert\lambda_2\vert>1,$ 
	and $\lambda_{2}=\overline{\lambda}_1\in{\rm i}\mathbb{R},$ $|\lambda_1|=1.$ 
In the second case all solutions (\ref{eq:solForm}) are non-decaying oscillatory and therefore irrelevant to our study. In the first case 
$\vU$ is a linear combination of exponentials decaying at $-\infty$ or $\infty.$ The condition of decay as $x\rightarrow\infty$ implies that 
\begin{equation} 
\label{eq:DecayCond}
\vU(x)
		={c}_{1}
		\exp\left(\dfrac{x}{d}\ln\lambda_{1}\right)\widetilde{\Psi}_{1}(x),
		\quad x>0.
\end{equation}

Imposing a specific boundary condition\footnote{As we discuss in Section \ref{Classical_Leontovich_cond_sec}, it is customary to impose a Leontovich condition 
$(E_1, E_2)^\top= Z(\mathbf{n} \wedge (H_1, H_2)^\top)$ on the boundary, 
 see \cite{BabichKuznetsov}, \cite{Senior}. Alternatives include the ``metallic'' condition $E_{3}=0,$ obtained by setting the impedance $Z$ to zero.} at $x=0$ thus links the two components of the vector $\widetilde{\Psi}_1(0)={\mathbb T}_1,$ {\it i.e.} the first column of $\mathbb T.$ This provides an equation for $k, \omega$ describing the set 
of pairs $(k, \omega)$ for which there is a surface wave satisfying the required boundary condition.

\section{Leontovich condition at the boundary of a stratified half-space} 
\label{GLC}

\subsection{Half-spaces in contact} \label{FullSpaceProb}

We consider the situation, see \Fref{FSDiagram}, where the half-space $\{x_{3}<0\}$ is occupied by a Lorentz material with $\omega$-dependent permittivity $\epsL$ as in \Eref{lorEps} and constant permeability $\mu$, while the complementary half-space $\{x_{3}\ge 0\}$ is occupied by a stratified dielectric as described in Section \ref{Maxwell_section}. 
\begin{figure} [h]
\center
  \includegraphics[width=0.82\textwidth]{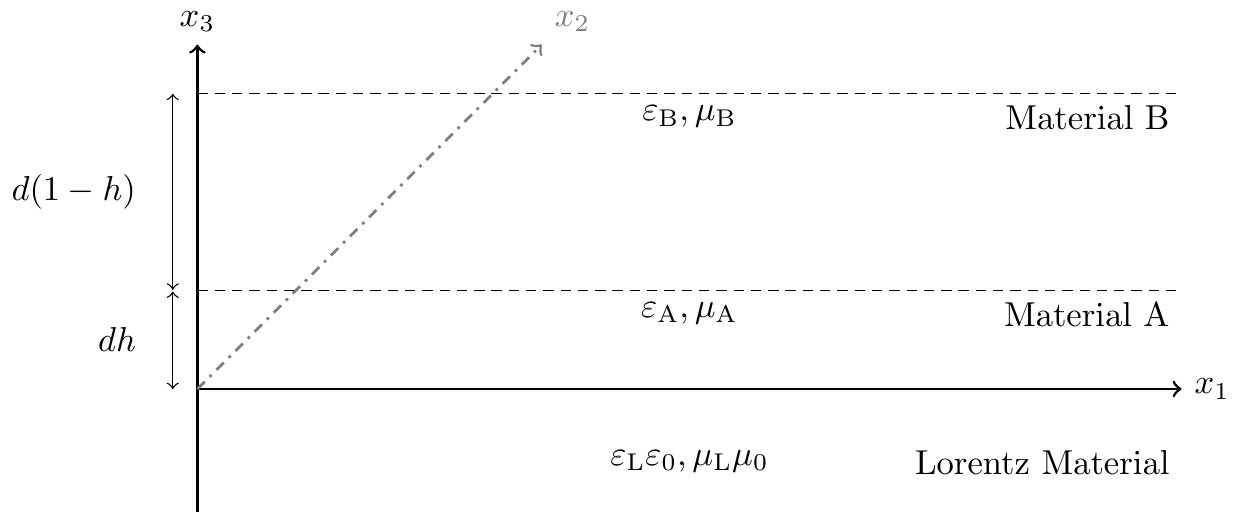}
  \caption{Diagram of the full-space Maxwell system. \label{fig:FSDiagram}} 
\end{figure}
We study the Maxwell problem in the entire space where the coefficients
$\varepsilon, \mu$ in \Eref{Maxwell} take the values corresponding to the material occupying each region of the space and seek interfacial wave solutions of the form \Eref{SolForm}. 
At each interface we impose the standard conditions that the quantities $\vE \cdot \mathbf{n}$, $\vH \cdot \mathbf{n}$, $\vD \wedge \mathbf{n}$, $\vB \wedge \mathbf{n}$ be continuous, where $\mathbf{n}$ denotes the normal the interface. 
Seeking a wave on the interface between the two media, we impose the condition of (exponential) decay away from the plane $\{x_{3}=0\}.$
We focus on the TE polarisation, when $\vU=(E_1, H_2)^\top$. 
We solve the half-space problem in each of the two media and couple the solutions at the shared boundary $\{x_{3}=0\}.$

\subsection{Equivalent problem in a single half-space, and dispersion relations} \label{sec:EqHSProb}

Having obtained the solution in the half-space $\{x_3<0\},$ see Section
\ref{sec:LowLor}, we find that the full-space problem supports an interfacial wave when the fields are related at $\{x_{3}=0\}$ via 
\Eref{GeneralLeontovich} whenever $\Re(\aL)>0$. 
The condition  \Eref{GeneralLeontovich}  captures the effect of the Lorentz material on the full-space system: we may solve the equivalent half-space problem for a stratified dielectric, using (\ref{eq:GeneralLeontovich})
as a boundary condition 
and seeking surface wave solutions in the stratified half-space propagating along $\{x_{3}=0\}$. 

We apply the boundary condition \Eref{GeneralLeontovich} to the solution (\ref{eq:DecayCond}) in the half-space $\{x_3>0\}$ obtained in Section \ref{Maxwell_section}. The corresponding monodromy matrix, see Section \ref{Floquet_section}, is given by 
\[
{\mathbb T}=\left(\begin{array}{cc}C_{\rm B}C_{\rm A}+\dfrac{\chi_{\rm B}\varepsilon_{\rm A}}{\chi_{\rm A}\varepsilon_{\rm B}}S_{\rm B}S_{\rm A}&-\dfrac{{\rm i}\chi_{\rm A}}{\varepsilon_{\rm A}}C_{\rm B}S_{\rm A}-\dfrac{{\rm i}\chi_{\rm B}}{\varepsilon_{\rm B}}S_{\rm B}C_{\rm A}\\[1.6em]
\dfrac{{\rm i}\varepsilon_{\rm A}}{\chi_{\rm A}}C_{\rm B}S_{\rm A}+\dfrac{{\rm i}\varepsilon_{\rm B}}{\chi_{\rm B}}S_{\rm B}C_{\rm A}&C_{\rm B}C_{\rm A}+\dfrac{\chi_{\rm A}\varepsilon_{\rm B}}{\chi_{\rm B}\varepsilon_{\rm A}}S_{\rm B}S_{\rm A}
\end{array}\right),
\]
where we use the following expressions involving the dimensionless wavenumber $\widehat{k}=dk$ and phase velocity ${v_{\rm p}}=\omega/k:$
\begin{equation*}
\begin{aligned}
\chiA&:= \sqrt{1-{v_{\rm p}^2}\muA\epsA}, 
\quad \chiB = \sqrt{1-{v_{\rm p}^2}\muB\epsB},\\[0.4em]
S_{\rm A}&:= \sinh(\chiA\widehat{k}h), 	\quad S_{\rm B}:= \sinh\bigl(\chiB\widehat{k}(1-h)\bigr), \quad
C_{\rm A}:= \cosh(\chiA\widehat{k}h), \quad C_{\rm B}: = \cosh\bigl(\chiB\widehat{k}(1-h)\bigr). 
\end{aligned}
\end{equation*}
Following the argument of Section \ref{sec:UppHSSol}, we are interested in the values of $k$ and $\omega$ for which  
$(-{\rm i}\alpha_{\rm L}/(\omega\varepsilon_{\rm L}\varepsilon_0), 1)^\top$ is an eigenvector of the matrix ${\mathbb T}$ with an eigenvalue whose absolute value is smaller than one. As a result, we obtain the dispersion relation for interfacial wave solutions:
\begin{align}
\label{eq:DispRelFullGeneral}
&\left[ \frac{\chiA\epsB}{\chiB\epsA} - \frac{\chiB\epsA}{\chiA\epsB} \right]{S}_{\rm A}{S}_{\rm B} + \left[\frac{\chiL\epsA}{\epsL\epsO\chiA} - \frac{\epsL\epsO\chiA}{\chiL\epsA}\right]{S}_{\rm A}{C}_{\rm B} 
+ \left[\frac{\chiL\epsB}{\epsL\epsO\chiB}- \frac{\epsL\epsO\chiB}{\chiL\epsB}\right]{S}_{\rm B}{C}_{\rm A}=0,\\[0.5em]
&\chiL:= \sqrt{1-{v_{\rm p}^2}\muO\muL\epsL\epsO}.\nonumber
\end{align}
The admissible $(\widehat{k},\vp)$ 
must further satisfy the conditions $\Re(\aL)>0,$ 
where $\aL$ is given by (\ref{alpha_L}), and 
\begin{equation}
\biggl\vert\biggl(\dfrac{\varepsilon_{\rm B}}{\chi_{\rm B}}S_{\rm B}C_{\rm A}+\dfrac{\varepsilon_{\rm A}}{\chi_{\rm A}}C_{\rm B}S_{\rm A}\biggr)\frac{\chi_{\rm L}}{\varepsilon_{\rm L}\varepsilon_0}+\dfrac{\chi_{\rm A}\varepsilon_{\rm B}}{\chi_{\rm B}\varepsilon_{\rm A}}S_{\rm B}S_{\rm A}+C_{\rm B}C_{\rm A}\biggr\vert<1.
\label{lambda_cond}
\end{equation}
The TM polarisation gives a dispersion relation similar to the above, see 
a discussion at the end of Section
\ref{sec:SingLayMax}.


The condition \Eref{DispRelFullGeneral} in general provides two constraints on $\widehat{k}$ and ${v_{\rm p}},$ namely that the real and imaginary parts of the expression vanish. With the exception of a case of homogeneous dielectric occupying 
$\{x_{3}>0\},$ the real part in \Eref{DispRelFullGeneral} does not vanish identically in any region of $(\gamma, \omega, k).$ Should this happen to the imaginary part, 
one obtains dispersion branches provided by a single equation\footnote{We can interchange between the pairs $(k,\omega)$ and $(\widehat{k},\vp)$ freely, so a dispersion relation between one pair gives an equivalent one between the other.} in $(k,\omega)$ for a fixed value of $\gamma.$ One such situation 
is if the Lorentz material is lossless ($\gamma=0$), when one obtains an explicit dispersion relation possessing multiple dispersion branches, see Section \ref{LosslessSystem}. 

\begin{figure} [h!]
\center
  \includegraphics[width=0.82\textwidth]{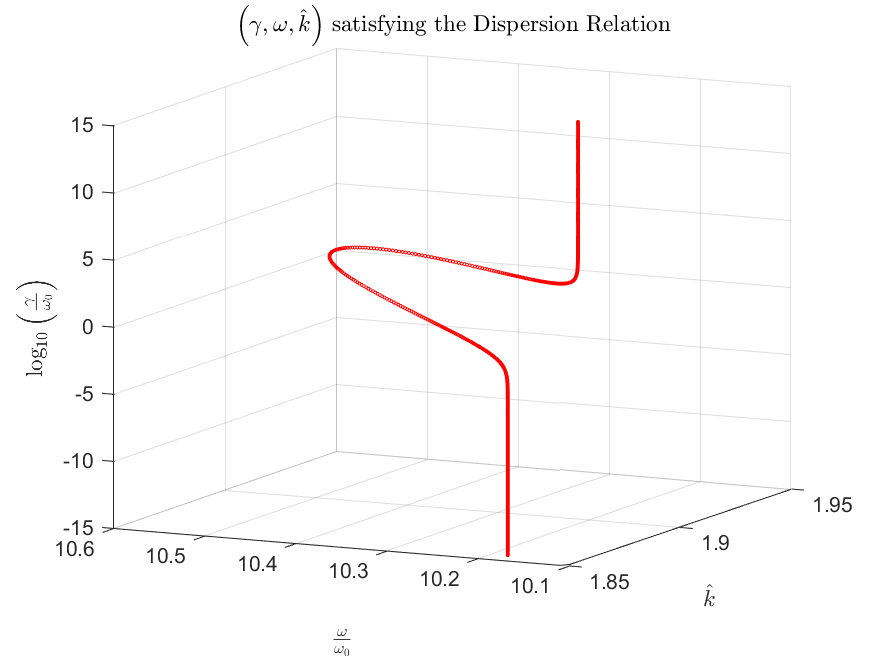}
  \caption{{\sc Solutions $(\gamma, \omega, k)$ to \Eref{DispRelFullGeneral}.} The plot shows the dependence of $(\omega, k)$  on the values of $\gamma$ such that $\log_{10}(\gamma/\omega_0)\in[-15,15].$ Parameter values used:  $h=0.5$, $\muA=\muB=\muO$, $\epsA=5\epsO$, $\epsB=10\epsO,$ $\mu_{\rm L}=1.$ The value for $\wP/\wO=2.13$ is the same as in   \cite{AlmogEtAl}.
  \label{3Dfigure}}
  \end{figure}

In general, the imaginary part in \Eref{DispRelFullGeneral}
is a function of $\gamma, \omega, k:$ 
one can express (at least locally)  
 $\gamma$ in terms of $\omega,$ $k$ and substitute it into 
the real part of \Eref{DispRelFullGeneral}, to obtain a dispersion relation in $k$ and $\omega$ only. 
We do not pursue this approach analytically, however an example of a numerical solution $(\gamma, \omega, k)$ to \Eref{DispRelFullGeneral} is provided in Figure \ref{3Dfigure}.
\begin{figure} [h!]
  \centering  
  \subfloat[Projection of the $(\gamma,\omega,\hat{k})$ onto the $(\gamma, \hat{k})$-plane.]{\includegraphics[width=0.45\textwidth]{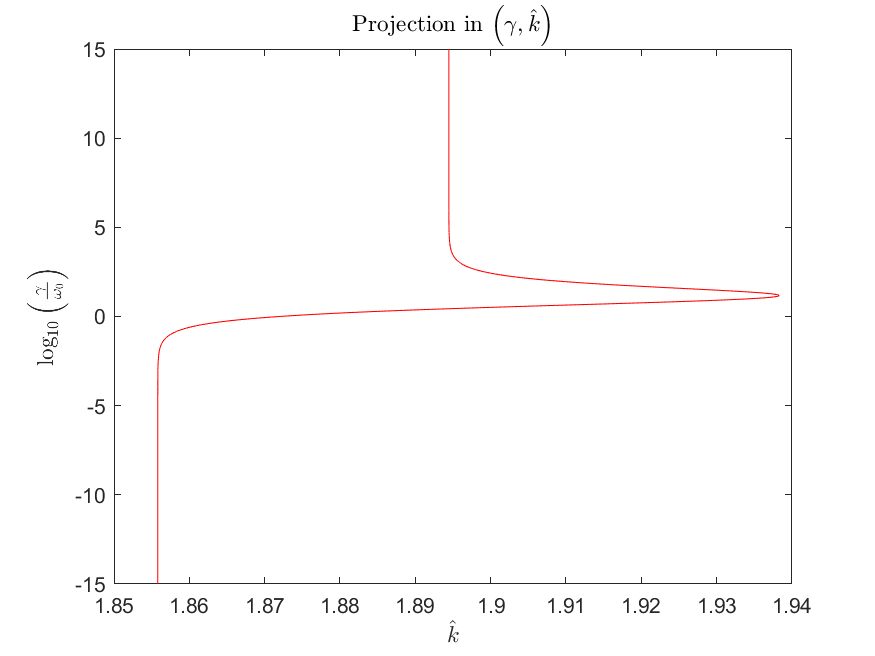}}
   \hfill
  \subfloat[Projection of the $(\gamma,\omega,\hat{k})$ onto the $(\omega, \hat{k})$-plane.]{\includegraphics[width=0.45\textwidth]{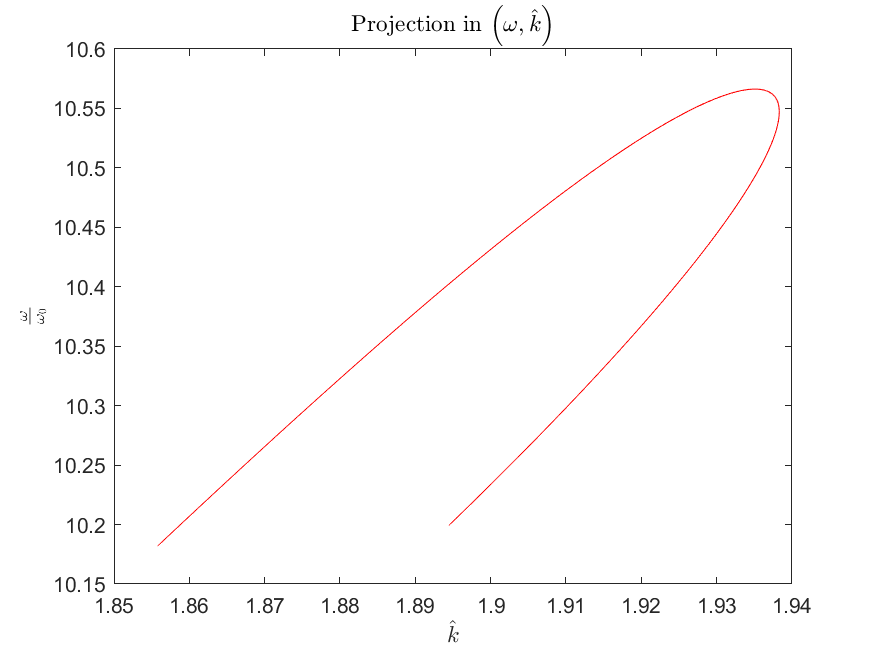}}
 \hfill
  \subfloat[Projection of the $(\gamma,\omega,\hat{k})$ onto the $(\gamma, \omega)$-plane.]{\includegraphics[width=0.45\textwidth]{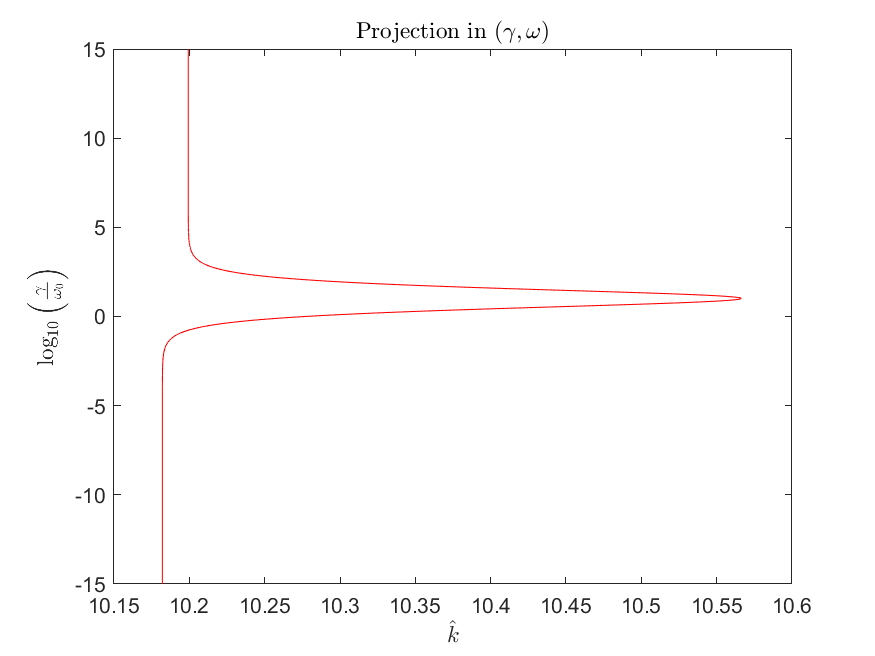}}
  \caption{{\sc Solutions $(\gamma, \omega, k)$ to \Eref{DispRelFullGeneral} in projections.} The plots show the projections of the curve in Figure \ref{3Dfigure} onto different coordinate planes in the $(\gamma, \omega, k)$-space.
 The range for $\gamma$ and the parameter values are the same as in Figure \ref{3Dfigure}.}
\end{figure}
\begin{remark}
We have so far assumed the magnetic permeability $\muL$ of the Lorentz medium to be a fixed material constant. However, it can be more generally modelled as $\omega$-dependent, when
takes a form analogous to $\epsL$ in \eqref{eL_er}--\eqref{eq:lorEps} and is treated as a function of the frequency $\omega,$ see Section \ref{Lorentz_section}.
\end{remark}


\subsection{Interfacial waves for a lossless Lorentz medium and stratified dielectric} \label{LosslessSystem}

In the case of a lossless Lorentz material
$\varepsilon_{\rm L}=\varepsilon_{\rm r}$, 
 {\it cf.} (\ref{eL_er})--(\ref{eq:lorEps}).
As before, the pairs $(\widehat{k},\vp)$ that satisfy (\ref{eq:DispRelFullGeneral}) are also subject to the conditions $\Re(\alpha_{\rm L})>0$ and (\ref{lambda_cond}).
The corresponding dispersion branches on the region $\omega\le50\omega_0,$ $dk\le10$ are shown in \Fref{LLDWOKall}. Notably, lowest branch possesses cut-on values for the frequency and wave-number $\omega/\omega_0=1,$ $\hat{k}=0.526,$ below which no interfacial waves exist. 


\begin{figure}[!h]
  \centering
 \subfloat[ The value $\wP/\wO=2.13$ is the same as in Figure \ref{3Dfigure}.
  ]
  {\includegraphics[width=0.45\textwidth]{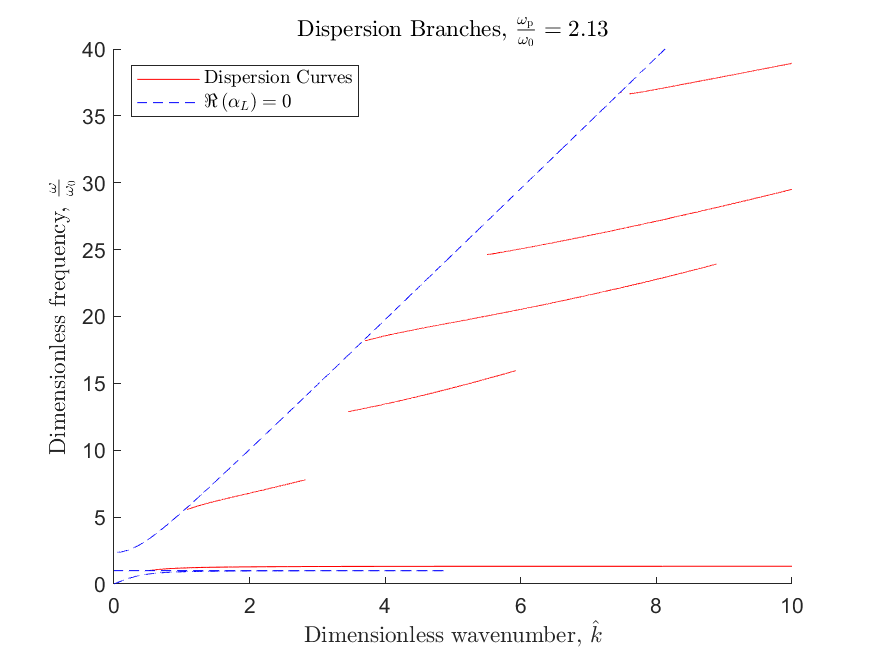}}
 \hfill
  \subfloat[The ratio of the plasma and resonant frequencies is increased to $\wP/\wO=5.$ 
  ]
   {\includegraphics[width=0.45\textwidth]{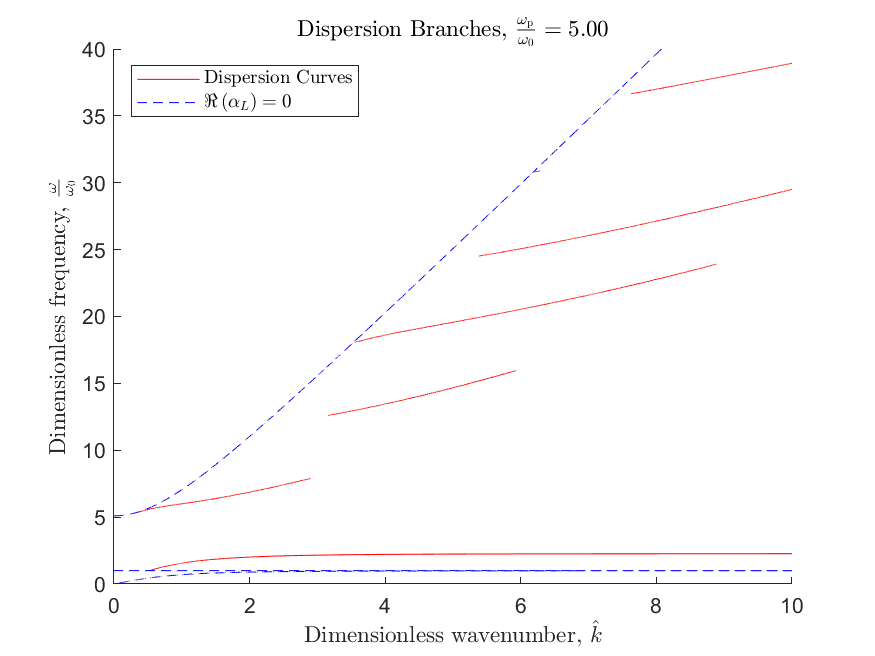}}
 \hfill
 \subfloat[The ratio of the plasma and resonant frequencies is increased further to $\wP/\wO=10.$ Long-wave solution appears at finite frequency.
  ]
  {\includegraphics[width=0.45\textwidth]{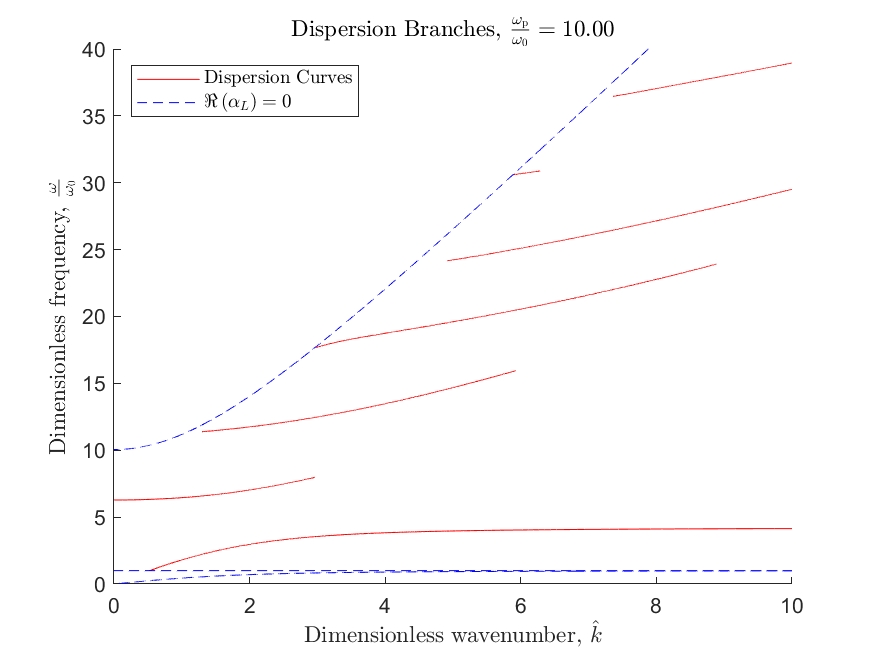}}
 \hfill
 \subfloat[The ratio of the plasma and resonant frequencies is increased further to $\wP/\wO=25.$ Long-wave solution appears at a larger number of finite frequencies.
  ]
  {\includegraphics[width=0.45\textwidth]{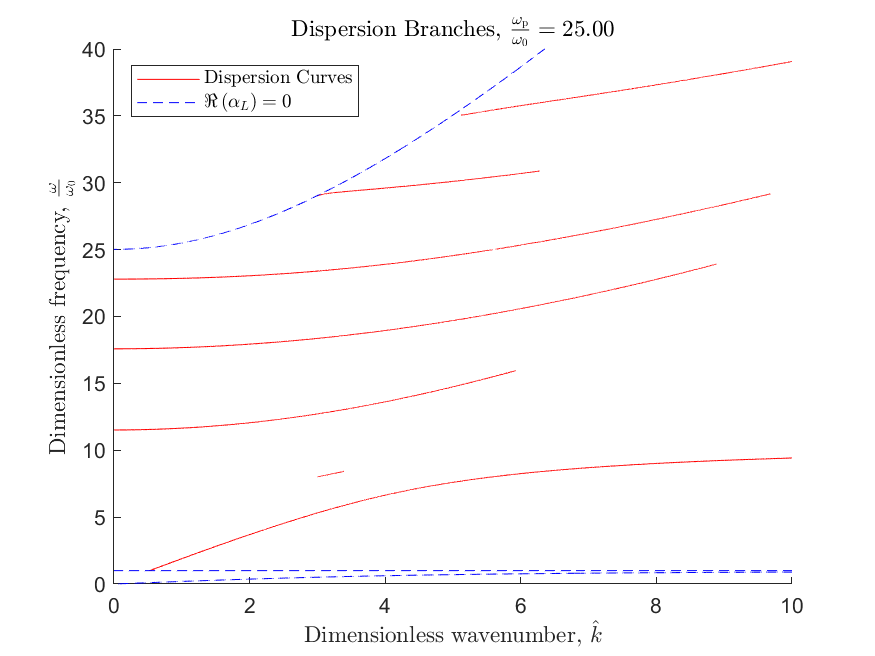}}
  \caption{{\sc Dispersion for lossless Lorentz half-space.} Dispersion branches that support decaying surface waves, when a lossless Lorentz material occupies the lower half-space. Parameter values used in each panel: $h=0.5$, $\muA=\muB=\muO$, $\epsA=5\epsO$, $\epsB=10\epsO,$ $\mu_{\rm L}=0,$ $\gamma=0.$  
    \label{fig:LLDWOKall}}
\end{figure}

\section{Half-space impedance condition} 
\label{GeneralToClassical}

\subsection{Classical Leontovich condition as a limit of (\ref{eq:GeneralLeontovich})}
\label{Classical_Leontovich_cond_sec}

It is common to make approximations to the exact interface conditions in systems similar to those in Section \ref{GLC}, by invoking a boundary condition at $\{x_{3}=0\}$ for one of the two half-spaces.
One such approximation is the classical Leontovich (or impedance) boundary condition, which requires that at the interface $\{x_{3}=0\}$, the tangential components of the $\vE$ field ($\vE_{\mathrm{t}}$) and the $\vH$  field  ($\vH_{\mathrm{t}}$) are related via
\begin{equation}
\label{eq:Leontovich}
\vE_{\mathrm{t}} = Z(\mathbf{n} \wedge \vH_{\mathrm{t}}). 
\end{equation}	
Here, $\mathbf{n}$ denotes the normal vector to the interface (pointing out from the material that is to be neglected, into the remaining material).
The quantity $Z$ represents an impedance, and more generally has the form $Z=\sqrt{\mu / \varepsilon}$ for a permittivity $\varepsilon$ and permeability $\mu$.
In this section we show that (\ref{eq:GeneralLeontovich}) can be used to recover the classical Leontovich condition. 


One can express the generalised impedance in (\ref{eq:GeneralLeontovich}) as follows:
\begin{align*}
-\frac{{\rm i}\aL}{\omega\epsL\epsO} 
= -\frac{\rm i}{\omega\epsL\epsO}\sqrt{k^{2}-\omega^{2}\muL\muO\epsL\epsO} 
= -{\rm i}\sqrt{\frac{\muL\muO}{\epsL\epsO}}\sqrt{\frac{k^{2}}{\omega^{2}\muL\muO\epsL\epsO}-1},
\end{align*}
in the case of TE polarisation, and
\begin{align*}
-\frac{{\rm i}\omega\muO\muL}{\aL} &= -{\rm i}\sqrt{\frac{\muO\muL}{\epsO\epsL}}\left(\frac{k^{2}}{\omega^{2}\muO\muL\epsO\epsL} -1 \right)^{-\frac{1}{2}}
\end{align*}
for TM polarisation.
For the case of constant $\muL$ and provided 
\begin{equation}
\abs{\frac{k^{2}}{\omega^{2}\epsL}}\ll 1,
\label{new_cod}
\end{equation}
we obtain the classical Leontovich condition with impedance $Z = \sqrt{\muO\muL/\epsO\epsL}$ as an approximation, up to the order 
$O\bigl(\bigl\vert \omega^{-2}\epsL^{-1}k^{2}\bigr\vert\bigr),$ 
to the condition
\Eref{GeneralLeontovich}.
To conclude, we note that in \cite{Senior} the Leontovich condition is purported to be derived under the condition that $\Mod{\epsL}\gg1,$ which coincides with (\ref{new_cod}) under the assumption that $\omega/k$ is bounded above and below.

\subsection{Homogeneous half-space with the classical Leontovich condition} \label{sec:BabichRecovery}

Under the assumption that the stratified dielectric half-space is actually homogeneous, \Eref{DispRelFullGeneral} is reduced to the equation
\begin{align*}
&\sinh\bigl(\chiA\widehat{k}\bigr)\left(\frac{\epsL\epsO\chiA}{\epsA\chiL} - \frac{\epsA\chiL}{\epsL\epsO\chiA}\right)=0,
\end{align*}
by setting\footnote{To obtain a homogeneous half space as a limit of the stratified system, one could also take the limit $h\rightarrow0.$} $\epsA=\epsB$ and $\muA=\muB.$
This equation has solutions for when either factor is zero, in the case of the $\sinh$ factor this happens only when $\chiA=0$ (and hence $\chiB=0$), which does not correspond to a decaying wave into the dielectric (the expression in (\ref{lambda_cond}) has modulus one, and the solution (\ref{eq:DecayCond})
has constant amplitude). 
In the case of the latter factor we may rearrange and obtain the dispersion relation
\begin{align*}
\omega^{2}	&= \frac{k^{2}}{\muA\epsO\epsL-\muO\muL\epsA}\left(\frac{\epsO\epsL}{\epsA}-\frac{\epsA}{\epsO\epsL}\right). \labelthis\label{eq:DispRelHomoHS2}
\end{align*}
Note that for a general Lorentz material \Eref{DispRelHomoHS2} still has non-zero real and imaginary part, due to the presence of $\epsL$. 
Therefore, the discussion of 
Section \ref{sec:EqHSProb} is still applicable here, although it is now possible to rearrange and obtain $\gamma$ as a function of $\omega$ and $k$. 


One can also obtain results concerning homogeneous dielectric systems with classical Leontovich boundary conditions from the more general system presented in Section \ref{GLC}, using the fact that the dispersion relation collapses to \Eref{DispRelHomoHS2}.
In the regime of bounded $k/\omega$ and with $\abs{\epsL}\gg 1$, the dispersion relation \Eref{DispRelHomoHS2} reduces, to leading order, to the relation obtained in \cite[Eq.\,(9)]{BabichKuznetsov}:
\begin{align*} 
\omega = \dfrac{kc}{\sqrt{\dfrac{\epsA\muA}{\epsO\muO}\left(1-\dfrac{\muO\muL}{\epsO\epsL}\dfrac{\epsA}{\muA}\right)}}.
\end{align*}

\section{Conclusions}
In our analysis of the full-space problem, we have demonstrated the possibility to obtain a dispersion relation as in \Eref{DispRelFullGeneral} that in general has non-trivial expressions for the real and imaginary part, thereby providing only finitely many points $(\omega, k)$ that support surface waves for each value of the loss parameter $\gamma.$ The imaginary component of the dispersion relation has the form $F(\gamma, \omega, k)=0$, and if it permits manipulation for $\gamma$ in terms of the frequency $\omega$ and wave-number $k$, one can obtain dispersion branches via substitution into the real part of the dispersion relation. Each point of such a dispersion relation corresponds to an individual value of the parameter $\gamma.$ In general, however, the equation $F(\gamma, \omega, k)=0$ is unlikely to admit a closed form for $\gamma$ as a function of $\omega,$ $k.$

In the case of a lossless Lorentz material (see section \ref{LosslessSystem}), {\it i.e.} when $\gamma=0,$ one can obtain dispersion branches $(\omega, k)$ that support (decaying) surface waves, see \Fref{LLDWOKall}. These exhibit two qualitative differences from the case of a stratified medium in contact with a non-dispersive dielectric (see Appendix), namely the presence of an additional low-frequency branch with a frequency cut-on at approximately the resonant frequency, as well as the presence of an increasing number of long-wave propagating modes for larger values of the ratio between the plasma and resonant frequencies.  


In the course of our analysis of the full-space Maxwell system, the interface conditions \Eref{GeneralLeontovich}
has been obtained for the Maxwell system \Eref{Maxwell} in the case when the lower (Lorentz) half-space has $\omega$-dependent permittivity as in \eqref{eL_er} and constant permeability.
This condition plays an analogous r\^{o}le to the classical Leontovich condition \Eref{Leontovich}, in that it allows one to reduce a full-space problem with an interface to a half-space problem, with a boundary condition derived from one of the constituent media.
The expression for the generalised impedance comes from the exterior Lorentz material, and other material parameters emerge from the stratified half-space to which the problem is reduced. In this sense, the approach can be viewed as a combination of the perspectives of \cite{Senior} and \cite{BabichKuznetsov}: in the former, the impedance boundary condition is derived for the Maxwell equations in what would be the analogue of our Lorentz half-space and in the latter these conditions were postulated in the complementary dielectric half-space. 

In conclusion, we note that the results of the present paper can be generalised to the context of linearised elasticity, non-local constitutive relations (such as those discussed in \cite{Chebakov_et_al}), as well as the case of a thin interfacial layer between heterogeneous and/or dispersive media. We postpone the related analysis to future publications. 


  
\section*{Appendix: Lorentz medium replaced by a non-dispersive dielectric}
\begin{figure} [h!]
\center
  \includegraphics[width=0.66\linewidth]{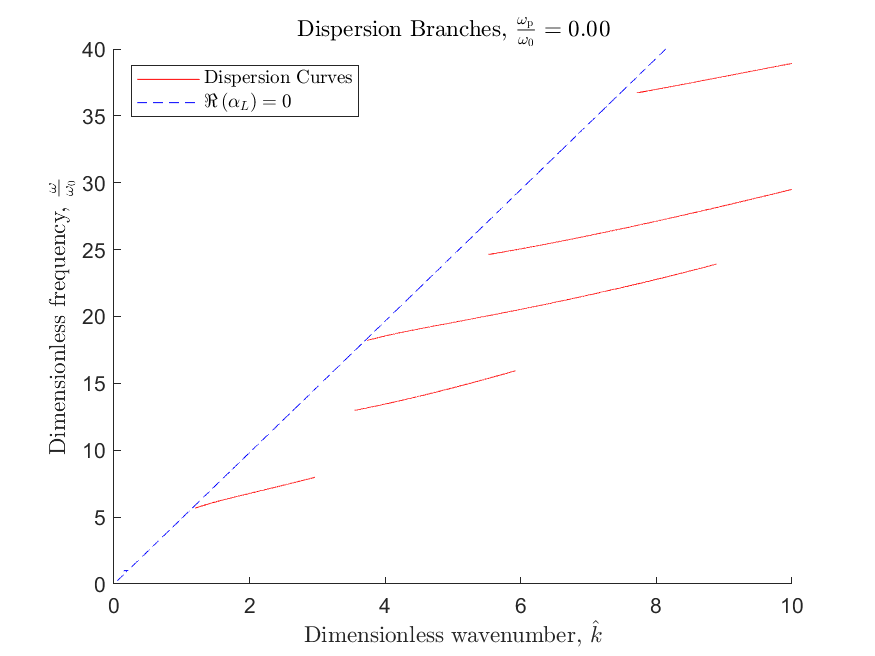}
  \caption{{\sc Dispersion curves for a homogeneous dielectric half-space in contact with a stratified dielectric.} 
  Parameter values used: $h=0.5, \muA=\muB=\mu_0,$ $\epsA=5\epsO,$ $\epsB=10\epsO,$ and $\varepsilon_{\rm L}=\mu_{\rm L}=1.$ The value $\wO=6.077\times10^{15}\,{\rm s}^{-1}$  is used to obtain the non-dimensional frequency $\omega/\omega_0.$  
  \label{fig:DispBranchesDSD_Bbigger}}
\end{figure}

When the half-space occupied by the Lorentz material is filled with a non-dispersive dielectric instead, we obtain dispersion curves shown in Figure \ref{fig:DispBranchesDSD_Bbigger}.

\begin{figure}[!h]
  \centering
 \subfloat[The ratio $\wP/\wO$ is set to unity. 
  ]
  {\includegraphics[width=0.45\textwidth]{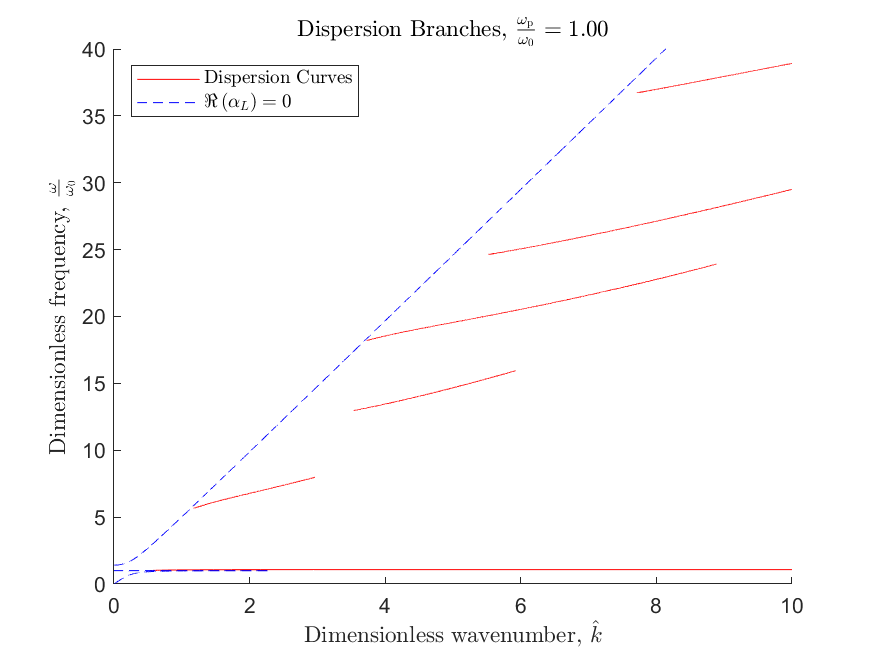}}
  \hfill
   \subfloat[Lowest dispersion branch. The ratio $\wP/\wO$ is set to unity, as in the panel (a). 
  ]
  {\includegraphics[width=0.45\textwidth]{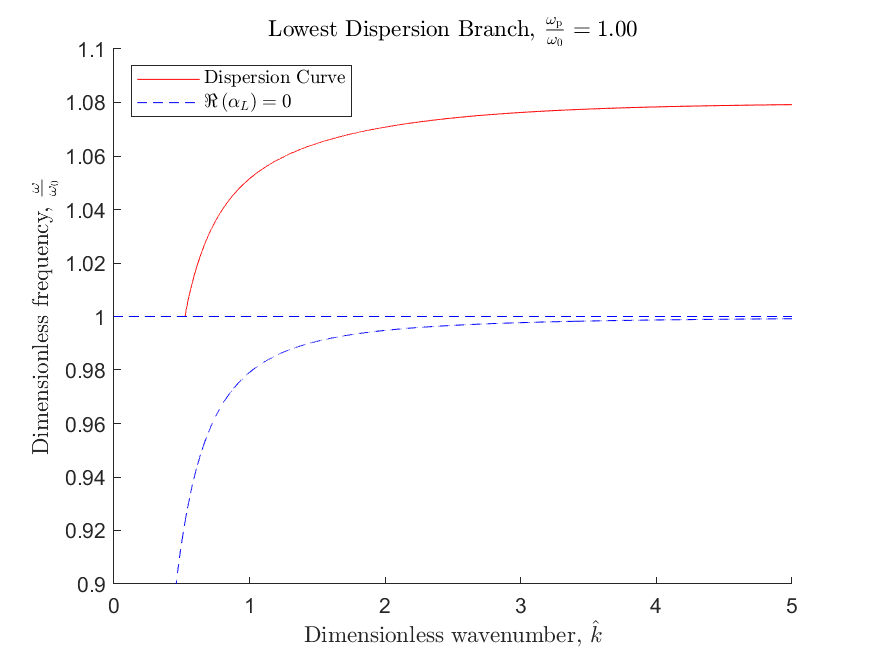}}
 \hfill
  \subfloat[Lowest dispersion branch. The ratio of the plasma and resonant frequencies is decreased to $\wP/\wO=0.1.$ 
  ]
   {\includegraphics[width=0.45\textwidth]{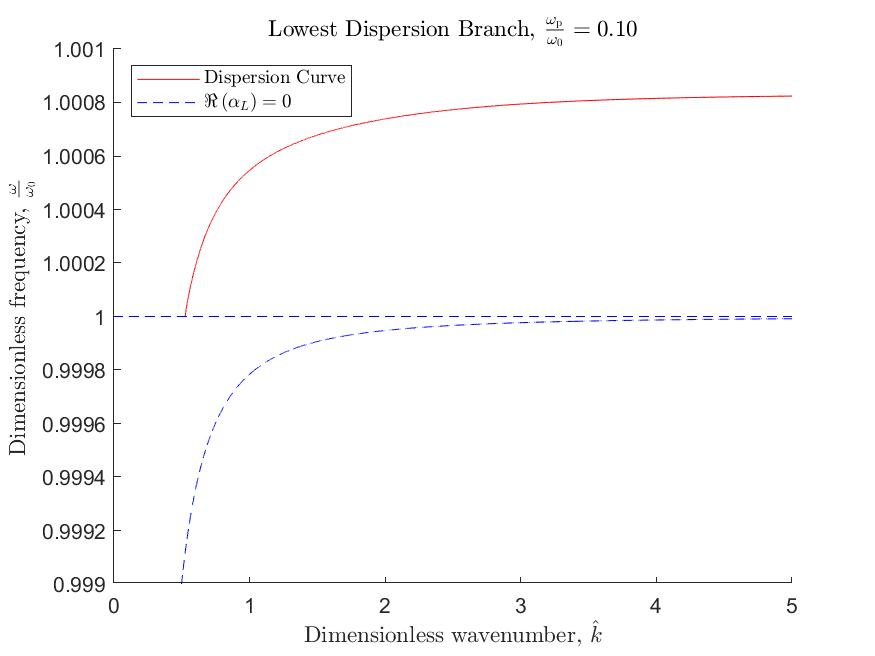}}
 \hfill
 \subfloat[Lowest dispersion branch. The ratio of the plasma and resonant frequencies is decreased further to $\wP/\wO=0.01.$
  ]
  {\includegraphics[width=0.45\textwidth]{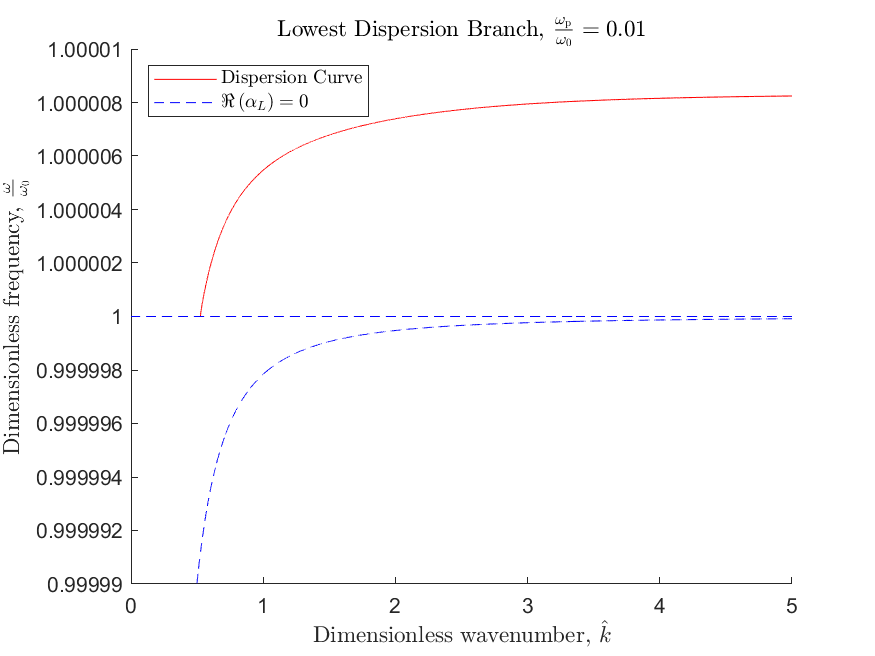}}
 \hfill

  \caption{{\sc Waves along the interface between a dispersive half-space and a stratified dielectric, as $\omega_{\rm p}/\omega_0\to0.$} 
  Parameter values used in each panel: $h=0.5,$ $\muA=\muB=\mu_0,$ $\epsA=5\epsO,$ $\epsB=10\epsO,$ $\mu_{\rm L}=1,$ and $\gamma=0.$ The value $\wO=6.077\times10^{15}\,{\rm s}^{-1}$  is used to obtain the non-dimensional frequency $\omega/\omega_0.$  
  \label{fig:DispBranchesDSD_Bbigger1}}
\end{figure}

The convergence, as $\omega_{\rm p}/\omega_0\to0,$ of the dispersion diagrams for waves along the interface between a lossless Lorentz dielectric half-space and a stratified dielectric is illustrated in Figure \ref{fig:DispBranchesDSD_Bbigger1}.

\section*{Acknowledgements}

KC is supported by Engineering and Physical Sciences Research Council: Grant EP/L018802/2 ``Mathematical foundations of metamaterials: homogenisation, dissipation and operator theory". WG is supported by a scholarship from the EPSRC Centre for Doctoral Training in Statistical Applied Mathematics at Bath (SAMBa), under the project EP/L015684/1.




\end{document}